\documentclass[journal]{IEEEtran}
\usepackage{cite}
\ifCLASSINFOpdf
  \usepackage[pdftex]{graphicx}
  \graphicspath{{img/}}
  \DeclareGraphicsExtensions{.pdf,.jpeg,.png}
\else
\fi
\usepackage{amsmath}
\usepackage{amssymb}
\usepackage{bm}
\usepackage{array}
\usepackage{hyperref}
\usepackage{booktabs}
\usepackage{eurosym}
\usepackage{xcolor}
\usepackage{balance}
\usepackage{fancyhdr}

\begin{document}
\title{The Importance of Context When Recommending TV Content: Dataset and Algorithms}
\author{Miklas~S.~Kristoffersen,
        Sven~E.~Shepstone,
        and~Zheng-Hua~Tan
\thanks{M.S. Kristoffersen is with the Research Department of Bang \& Olufsen A/S, 7600 Struer, Denmark, and the Signal and Information Processing Section, Department of Electronic Systems, Aalborg University,
9220 Aalborg~\O, Denmark (email: mko@bang-olufsen.dk).}
\thanks{S.E. Shepstone is with the Research Department of Bang \& Olufsen A/S, 7600
Struer, Denmark (e-mail: ssh@bang-olufsen.dk).}%
\thanks{Z-H. Tan is with the Signal and Information Processing Section, Department of Electronic Systems, Aalborg University,
9220 Aalborg~\O, Denmark (e-mail: zt@es.aau.dk).}%
\thanks{The work is supported by Bang \& Olufsen A/S, Denmark, and the Innovation Fund Denmark (IFD) under File No. 5189-00009B.}}%

\fancypagestyle{specialfooter}{%
    \fancyhf{}
    \renewcommand\headrulewidth{0pt}
    \fancyfoot[C]{\tiny Accepted for publication in IEEE Transactions on Multimedia -- Digital Object Identifier 10.1109/TMM.2019.2944214. \copyright 2019 IEEE. Personal use of this material is permitted. Permission from IEEE must be obtained for all other uses, in any current or future media, including reprinting/republishing this material for advertising or promotional purposes, creating new collective works, for resale or redistribution to servers or lists, or reuse of any copyrighted component of this work in other works.}
}

\maketitle

\begin{abstract}
Home entertainment systems feature in a variety of usage scenarios with one or more simultaneous users, for whom the complexity of choosing media to consume has increased rapidly over the last decade.
Users' decision processes are complex and highly influenced by contextual settings, but data supporting the development and evaluation of context-aware recommender systems are scarce.
In this paper we present a dataset of self-reported TV consumption enriched with contextual information of viewing situations.
We show how choice of genre associates with, among others, the number of present users and users' attention levels.
Furthermore, we evaluate the performance of predicting chosen genres given different configurations of contextual information, and compare the results to contextless predictions.
The results suggest that including contextual features in the prediction cause notable improvements, and both temporal and social context show significant contributions.
\end{abstract}

\IEEEpeerreviewmaketitle

\thispagestyle{specialfooter}

\section{Introduction}
The underlying factors affecting users' choices of what to watch on TV have for several years been of interest to commercial and academic research.
In the midst of a rapidly changing device and multimedia landscape, TVs continue to be at the core of multimedia consumption in the home with scenarios covering, among others, social gatherings and solitary immersive moments.
The inherent complexity of viewing situations challenges the creation of experiences that match personal preferences as well as
preferences introduced by the situations themselves.
That is, priorities stemming from aspects such as the temporal and social settings of a viewing situation, generally referred to as the context.

Due to the increased availability of multimedia, the focus of recent research has been on improving the users' decision process by reducing large catalogs of content to a few personalized suggestions \cite{Veras2015}.
Commercial recommender solutions are now considered core to the business of engaging users and thereby preventing abandonment \cite{Gomez-Uribe2015}. 
To do so, recommender systems have explored various features for personalization, such as history of watching, ratings, user/item similarity, and time of the day, the last of which is an example of features characteristic to context-aware recommender systems (CARS) \cite{Adomavicius2015}.
The main objective of traditional recommender systems is to personalize the experience to the individual, often by studying the user-item matrix.
This is an issue, since a TV is often shared by multiple members of a household that engage in both solitary and social viewing situations during everyday life.
A solution for accommodating different solitary users on one TV is to allow multiple accounts, but this does not solve the problem when a group of users want a shared experience~\cite{Gomez-Uribe2015}, which is the focus within the research field of group recommender systems~\cite{Masthoff2015}.
Hence, to achieve recommendations aimed at dynamic compositions of users, the system needs to be aware of the social context it is used in, and thus more features than those classically used for personal predictions are needed~\cite{Shi2014,Cui2016}.
These features can be in the form of personal preferences for each member of the group as seen in~\cite{Masthoff2015}, e.g. based on social attributes aggregated from social media~\cite{Cui2014,Bertini2013}, or a more general social context as seen in~\cite{Cremonesi2015}, e.g. a group of adults watching TV.
In this work, we adopt the latter approach and describe social situations with e.g. the number of viewers and their relationship.
Recent studies even suggest to decouple the goal of tailoring the experience to the individual (personalize) from tailoring to the situation and intent (contextualize) \cite{Pagano2016}, thereby focusing more on the immediate context than the past behavior of a user.
In this study, we investigate some of the advantages and disadvantages of contextless personalization as well as contextualized suggestions. We focus less on algorithmic improvements within each approach, e.g. optimization of contextless personalization, and more on the contribution of different contextual dimensions.

Even though the concept of context-aware recommendations has been studied in several academic and commercial projects, there is still a need for publicly available datasets since only a limited number of such datasets exist, e.g. \cite{Turrin2014}.
Furthermore, the majority of existing CARS datasets are based on explicit feedback, often in the form of ratings.
However, within TV recommender systems the feedback is usually implicit in the form of watched/not watched, since continuously probing users for explicit feedback would significantly alter the viewing experience.
The use of implicit feedback also means that a user can provide feedback for the same content multiple times, which is typically not the case for e.g. movie ratings.

Another challenge of using existing datasets for developing CARS within the TV domain, is that the most dominant contextual feature is timestamps, i.e. temporal information.
Though TV viewing is highly driven by habits linked to temporal context such as time of day, day of week, and season, CARS based exclusively on temporal information could miss non-trivial correlations between e.g. temporal and social contexts.
It is, however, challenging to collect TV consumption data that include contextual information beyond timestamps.
People meters\footnote{A device used to log users' TV viewing behavior. For identification, the device often relies on participants to push a button on a remote when they enter and leave.}, for instance, are challenged, \cite{Jardine2016}, by non-compliance (participants neglect to push a button), and secondly, since meters can only log what was shown, and not necessarily whether the participant was engaged, there is no information of participants' actual exposure to content.
In other words, the TV could be showing some content that the participant does not watch.

In this study, we collect and analyze self-reported interactions for a limited number of content classes (genres).
We include several contextual features, which allow inspection of patterns in the consumption of different genres.
In addition to the comprehensively studied temporal information, our contribution includes novel investigations of associations between consumed content and social settings, e.g. who is present.
Using the Experience-Sampling Method (ESM) \cite{Larson1983}, we ask participants to report TV consumption multiple times each day for a five week period.
Through self-reported data, we decrease uncertainty of exposure to content, and allow collection of non-trivial information, such as how much attention is paid to the TV.
The data is structured to accommodate quantitative analyses, e.g.\ in the CARS community, and is publicly available under the name Contextual TV (CTV) dataset\footnote{Available at \url{http://kom.aau.dk/~zt/online/ContextualTVDataset}.}.
Note that the self-reported information provided by participants could potentially be collected implicitly at run-time in a real-world implementation using technologies such as user detection~\cite{Nguyen2016}, identification~\cite{Schroff2015}, attention level estimation~\cite{Coifman2019}, and indoor localization~\cite{Xiao2016}.

Using different feature configurations, we also show how well-established methods perform in predicting consumed content given contextual settings, and compare this with contextless prediction.
In an initial study, we showed the effectiveness of including contextual information \cite{Kristoffersen2018}, which we expand in the present contribution with in-depth data analysis and detailed investigation of prediction performance.
That is, we assess gains of adding contextual knowledge to the prediction task, and study contributions from each contextual dimension to the overall ability to predict which genre a user will engage with in a given situation.

The rest of the paper is organized as follows.
We start by surveying related work in Section~\ref{sec:related}.
In Section~\ref{sec:data} we introduce and analyze the contextual TV dataset.
Section~\ref{sec:pred} presents methods for predicting content and evaluates different configurations.
Finally, Section~\ref{sec:disc} and \ref{sec:conc} discuss the findings and conclude the study. 

\section{Related Work}\label{sec:related}
\subsection{Contextual Aspects of Watching TV}
Previous studies of users' TV watching behavior in given contexts have shown that the TV is mostly a social platform and consumption takes place in a wide variety of scenarios.
In \cite{Saxbe2011}, 30 households were scan-sampled every 10 minutes for four days, to reveal patterns of who was watching, when, and with whom.
Noticeably it was found that 64\% of the time, family members were watching TV together.
\cite{Abreu2013} presents the results of an online survey with 550 valid responses.
Their results confirm that contextual settings, and not only the users' personal profiles, are of importance to the decision of what content to watch.
In \cite{Mercer2014} a multi-method field study is conducted with 11 participants.
Both temporal and social settings are highlighted as key contextual indicators of consumed content.
An example of collecting qualitative contextual TV consumption data using diaries is presented in \cite{Vanattenhoven2015} for 12 households over a three week period.
The main difference between the studies listed above and our work, is that we aim for a quantitative dataset that avoids the recall bias associated with questionnaires.
To this end, ESM has proven useful for obtaining frequency and patterning of daily activities and social interactions~\cite{Csikszentmihalyi2014}.
In a recent study, \cite{Kim2018} combined automated data logging from the TV with event-triggered ESM to show, among others, how social context affects TV volume.
That is, they used a number of sensors to automatically extract contextual settings of TV viewing events, e.g. Bluetooth trackers to identify present users and their activity level, together with chatbot sessions for obtaining self-reported information of e.g. social context.

\subsection{Recommendations Based on Context}
The task of recommending content to users based on their past behavior as well as context, is an active research field.
Early work focused on pre- and post-filtering \cite{Adomavicius2015}, while recent studies have included contextual information directly in the model.
The main approaches are tensor factorization \cite{Frolov2017}, factorization machines \cite{Rendle2010,Rendle2011}, and most recently efforts based on deep learning such as \cite{Cheng2016}.
In \cite{Unger2017} an Android application is used to collect smartphone sensor-based contextual point of interest data from 90 students for a month long period, and recommendations are based on deep auto-encoding.
As in this work, they use ESM for collecting feedback from participants.
\cite{Feng2019} proposes a large-scale context-aware video retrieval method, and \cite{Benini2011} presents a method for recommending movies based on emotional preferences of users.

\begin{table}[tb]%
\centering%
\caption{Related Context-Aware TV Content Recommender Studies}%
\label{tab:related}%
\begin{tabular}{cccccc}%
\toprule%
    \textbf{Ref.} & \textbf{Temp.} & \textbf{Social} & \textbf{Mood/} & \textbf{Att.} & \textbf{Dataset}\\
    & & & \textbf{Emotion} & & \textbf{pub. avail.}\\\midrule
    \cite{Ardissono2004,Aharon2015,Bambia2016,Park2017} & \checkmark & $-$        & $-$         & $-$        & $-$\\
    \cite{Turrin2014}       & \checkmark & $-$        & $-$         & $-$        & \checkmark\\
    \cite{Shepstone2014}    & $-$        & $-$        & \checkmark  & $-$        & $-$\\
    \cite{Hsu2007}          & \checkmark & $-$        & \checkmark  & $-$        & $-$\\
    \cite{Vildjiounaite2009,Song2012,Cremonesi2015}& \checkmark & \checkmark & $-$         & $-$        & $-$\\
    CTV (ours)     & \checkmark & \checkmark & $-$         & \checkmark & \checkmark\\
\bottomrule%
\end{tabular}%
\end{table}%

Within TV content recommendation, several studies have based recommendations on context.
Table~\ref{tab:related} summarizes related works, and shows which contextual features are included in each study. 
The table also lists the availability of data from each contribution (to the best of the authors' knowledge).
An early example that includes contextual information is presented in \cite{Ardissono2004} that collected people meter data from approximately 60 participants for one year.
Their system has three main components: explicit user modeling based on explicit feedback from users; stereotypical user modeling based on age, gender, etc.; dynamic user modeling based on implicit information inferred from users' viewing behavior in certain temporal contexts.
Recently, \cite{Aharon2015} and \cite{Park2017} similarly studied temporal aspects of recommending TV content.
In \cite{Hsu2007} approximately 100 participants provided diary data for two weeks. 
Additional to temporal information, the collected viewing contexts include three mood selections (happy, bored, and unhappy).
Feedforward neural networks are used for recommending content, and their results suggest that users' emotional state helped improve the performance.
\cite{Vildjiounaite2009} presents recommendations using support vector machines (SVM) for people meter data collected from 20 families in Finland during a five month period.
The viewing contexts consist of temporal and social (additional viewers) information.
Their results suggest that including social context makes a minor improvement, but that the improvement depends on family habits, i.e. the correlation between temporal and social settings in families' typical viewing behavior.
\cite{Song2012} also includes social context, using RFID tags to identify users.
They evaluate their system in a real-world implementation.
In \cite{Shepstone2014}, users' moods are used to improve navigation of programs available in the electronic program guide (EPG).
\cite{Turrin2014} presents a people meter dataset containing implicit viewing events with timestamps for a four-month duration.
They smooth temporal context and use distance between contextual settings to recommend TV programs.
\cite{Cremonesi2015} presents a comparable dataset, but includes \textit{familiar context}, that is, the additional users watching.
Their results suggest that temporal context cancels the effect of social context when using both to recommend TV content.
In \cite{Bambia2016} more than 700 million views collected on a Tunesian TV platform are used to recommend TV content in a given context. 
Their viewing contexts are defined by location, time/day, weather and occasion. 

As evident from the literature listed above and in Table~\ref{tab:related}, context-aware TV recommendations have primarily revolved around quantitative data collected through (people) meters.
Though meters have a lot of advantages, such as enabling easy large-scale implicit feedback collection, they do (as previously stated) suffer from e.g. non-compliance and actual exposure uncertainty.
To better embrace the complexity of viewing situations, we ask participants to provide information that is not easily accessed through meters, such as how much attention a user is paying in a given viewing situation.
Also, instead of relying on participants to continuously register their presence in front of the TV using a remote control, our adopted data collection method is chosen to reduce noisy measurements of social settings.
Another observation from the literature is that studies have difficulties comparing their findings to those of other works, partially because there is no tradition of sharing results on common datasets, as is the tradition within e.g. movie recommendation.
Thus, to the best of the authors' knowledge, the present contribution is the first to publicly share a dataset with TV viewing events that include contextual settings beyond timestamps.

According to \cite{Lorenz2017}, users tend to choose content such that a few dominant items or providers account for the majority in consumption.
They refer to this phenomenon as contextual bias, and discuss how this decreases the diversity of recommenders.
In this paper, we include scores that enable assessment of diversity in predictions. 

\section{Contextual TV Watching Dataset}\label{sec:data}
This section details the procedure for collecting the dataset, and highlights a number of patterns within the data. 
The quantitative analysis is focused on general contextual tendencies of viewing situations as well as considerations of temporal and social context dynamics.

\subsection{Experimental Protocol}
\begin{table}[t]%
\centering%
\caption{Questions and Selection Options in the Dataset}%
\label{tab:ques}%
\begin{tabular}{m{0.03\columnwidth}m{0.39\columnwidth}m{0.42\columnwidth}}%
\toprule%
    \multicolumn{2}{l}{\textbf{Questions}} & \textbf{Options}\\\midrule
    Q1: & Have you watched TV within the last four hours? & Yes, no\\\midrule
    Q2: & Who were you watching it with? & \textit{Multiple-option:} Alone, partner, child~(0-12), child~(12+), sibling, parent, friend, other (text) \\\midrule
    Q3: & How many people (including yourself) watched TV? & 1, 2, 3, 4, 5+\\\midrule
    Q4: & What did you watch? & \textit{Multiple-option:} News, sport, movie, series, music, documentary, entertainment, children's, user-generated, other (text)\\\midrule
    Q5: & Which service(s) did you use? & \textit{Multiple-option:} Traditional TV, DRTV, TV2 Play, Viaplay, Netflix, HBO~Nordic, YouTube, other (text)\\\midrule
    Q6: & How much attention did you pay to the TV? & None-full (5 steps)\\%
\bottomrule%
\end{tabular}%
\end{table}%
To obtain data from participants we developed a web page, thereby allowing access from all devices equipped with Internet access and a web browser, though we recommend the use of mobile devices.
Participants were asked to answer questions five times every day at 8, 12, 17, 20, and 22 (or when going to bed) for a period of 36 days.
These intervals were chosen to accommodate work and study schedules, while  still providing ample opportunity to participate over a full day period.
Participants were allowed to answer more frequently and at other times than the five pre-specified intervals.
Scheduled reporting was preferred to event-based reporting, e.g.\ asking participants to answer after every TV watching session, since this requires less from the participants to remember and enables evaluation of compliance.
Also, scheduled reporting allows signaling to participants to help remind them about the study.
Specifically, we used a public calendar with alerts for iOS devices and web push notifications for all other types of devices.
Prior to launch, a three-day pilot test was conducted involving 12 potential participants evaluating web page and reminders.
Participants for the main study were recruited through social media with a lottery of three loudspeakers worth \euro 170 each as incentive at the end of the study.
A requirement for joining the lottery was at least 14 days of active participation.

The first time a user visits the web page that person is instructed to answer background information questions as part of the enrollment procedure. The collected information includes: Gender, age group, language (Danish/English), device type, household size, additional household members, frequency of TV watching, and favorite TV genres. 

On subsequent logins, participants were asked the questions listed in Table~\ref{tab:ques}.
The questions are designed to have a low cognitive load and take less than 30 seconds to answer.
The general flow is that Q2-Q6 are asked only if the selection for Q1 is \textit{yes}.
Also, Q3 is skipped if \textit{alone} is selected for Q2.
Note that in most cases it is not possible to infer the value of Q3 from Q2.
As an example, Q2 option \textit{friend} can be observed with all values for Q3 (except 1) if more friends are present.
For Q5 all except \textit{Traditional TV} (and possibly \textit{other}) are streaming services, some specific to Denmark/Scandinavia.
The multiple-option questions allow more than one selection, e.g. \textit{partner} and \textit{friend}.
Participants are instructed to split answers with different contextual settings, e.g.\ watching news alone and children's TV with a child.
Answers are logged with the following format:
Answer ID, User ID, timestamp, Q1, Q2, Q3, Q4, Q5, Q6.
In this study we extract two pieces of information from the timestamp.
One is the day of the week that can also be used to determine whether or not it is weekend.
Also, we group public holidays and weekend unless stated otherwise. 
The other feature is the time of the day.
We use five groups: 1) Morning: 6-10; 2) Noon: 10-14; 3) Afternoon: 14-18; 4) Evening: 18-22; 5) Night: 22-06.

\subsection{Data Analysis}\label{sec:data_analysis}
\subsubsection{Participants' Background}\label{sec:user_back}
A total of 118 participants (64 male and 54 female) in the age range 13-70 took part in the ESM study.
57\% of the participants were in the age range 21-30, and 84\% lived in households with at least two members including themselves.
At the enrollment, 81\% reported that they watched TV daily and 97\% that they watched at least once a week.
Concerning reported favorite genres, the \textit{series} genre attracts the largest audience~(89). 
\textit{Movie}~(75) and \textit{documentary}~(73) are also popular, while \textit{entertainment}~(63) and \textit{news}~(61) follow as fourth and fifth, respectively. 
\textit{Sport}~(44) is sixth, \textit{user-generated}~(21) seventh, and \textit{music}~(17) and \textit{children's}~(17) share the last place.

\subsubsection{Activity}
\begin{figure}[tb]%
    \centering%
    \includegraphics[width=1.0\columnwidth]{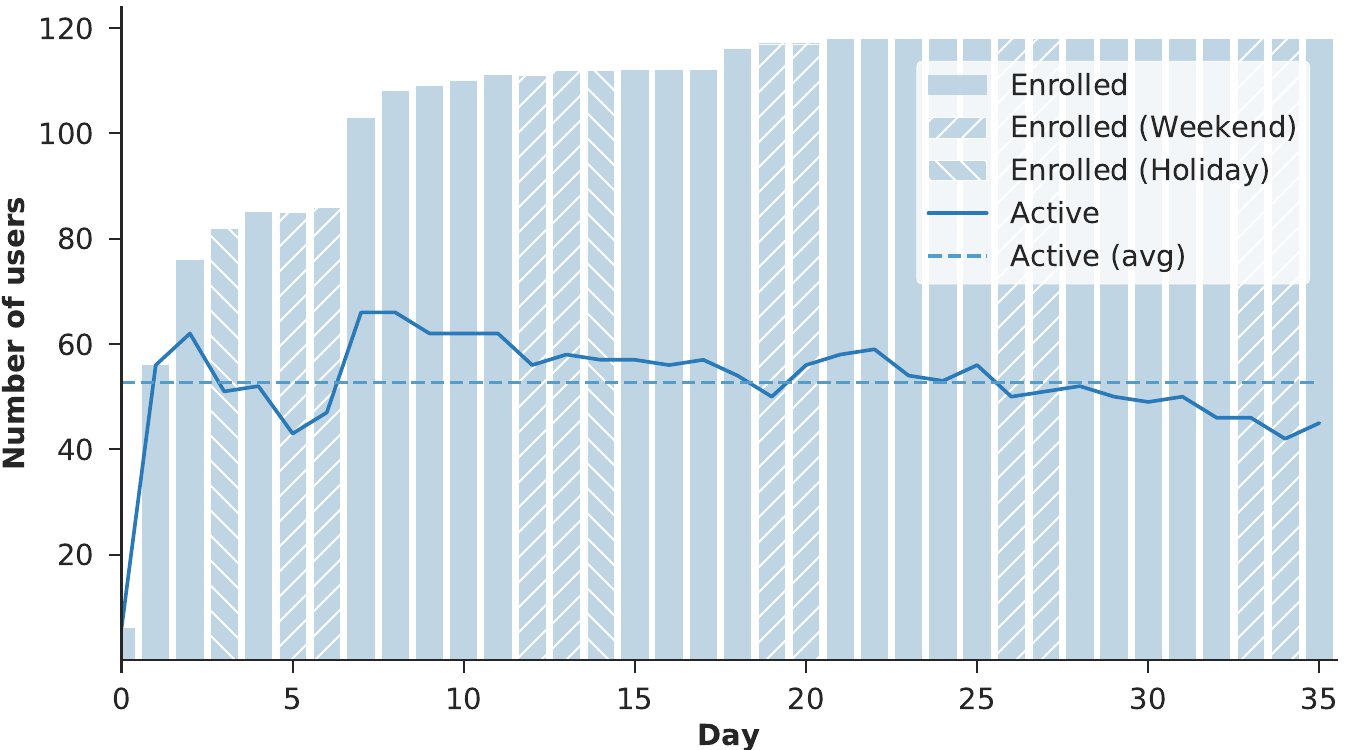}%
    \caption{Number of enrolled and active participants for each day of the study. The average number of active participants per day is shown with a dashed line.}%
\label{fig:fig1}%
\end{figure}

Fig.~\ref{fig:fig1} shows the development in enrolled and active (at least one answer that day) participants each day of the 36 days in the study. 
Notice the relatively large drop-off between enrolled and active participants within the first five days of the study, mainly caused by one-time visitors (a total of 31 throughout the study).
From day seven onwards, the number of active participants decreased on average by three every fourth day. 
The average number of active participants per day was 53, and a total of 60 participants met the requirement of at least 14 active days.
Saturdays (day 5, 12, 19, etc.\ of Fig.~\ref{fig:fig1}) had a tendency of fewer active participants, and also showed to be the day with the least responses (935) compared to the most on Wednesdays (1115). 
Notably, it seems that part of the inactive users returned Sundays after a "day off".
In terms of time of day, participants answered most frequently in the evening (1784), the least in the morning (1203), and the remaining ranked as follows: night (1468), noon (1418), and afternoon (1328).

\begin{figure}[tb]
    \centering
    \includegraphics[width=1.0\columnwidth]{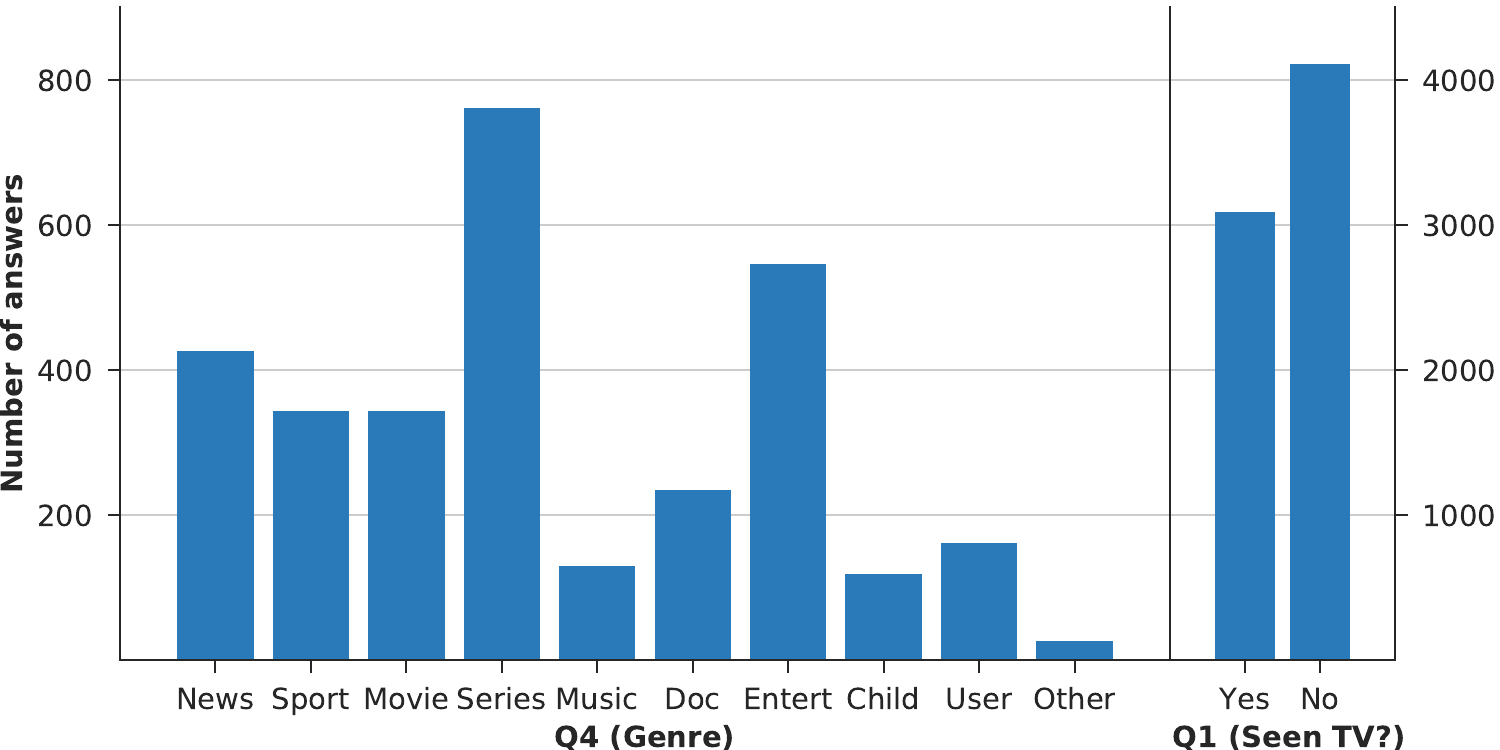}
    \caption{Count plots for answers to Q4 (left) and Q1 (right).}
\label{fig:fig2}
\end{figure}

\subsubsection{TV Consumption}
\begin{figure}[tb]%
    \centering%
    \includegraphics[width=1.0\columnwidth]{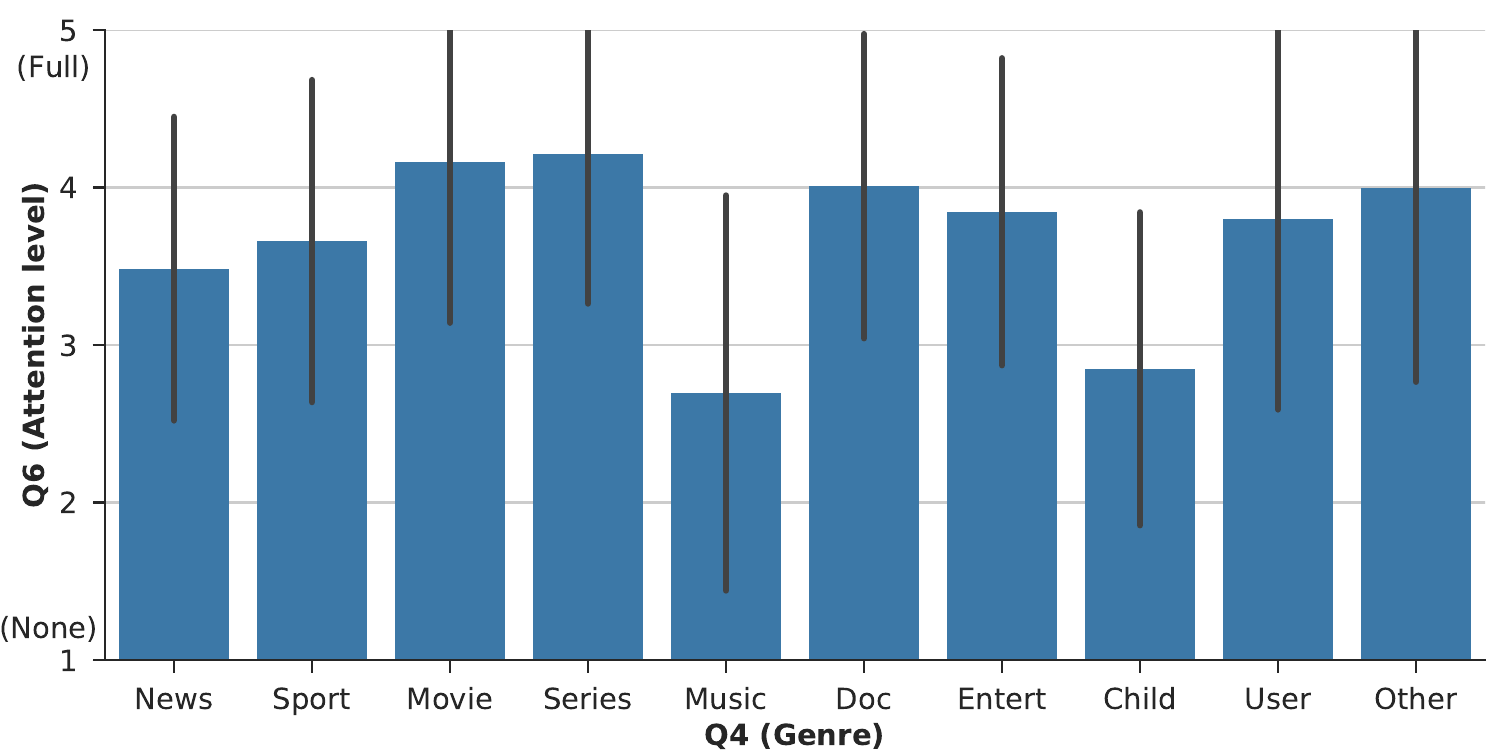}\\%
    \includegraphics[width=1.0\columnwidth]{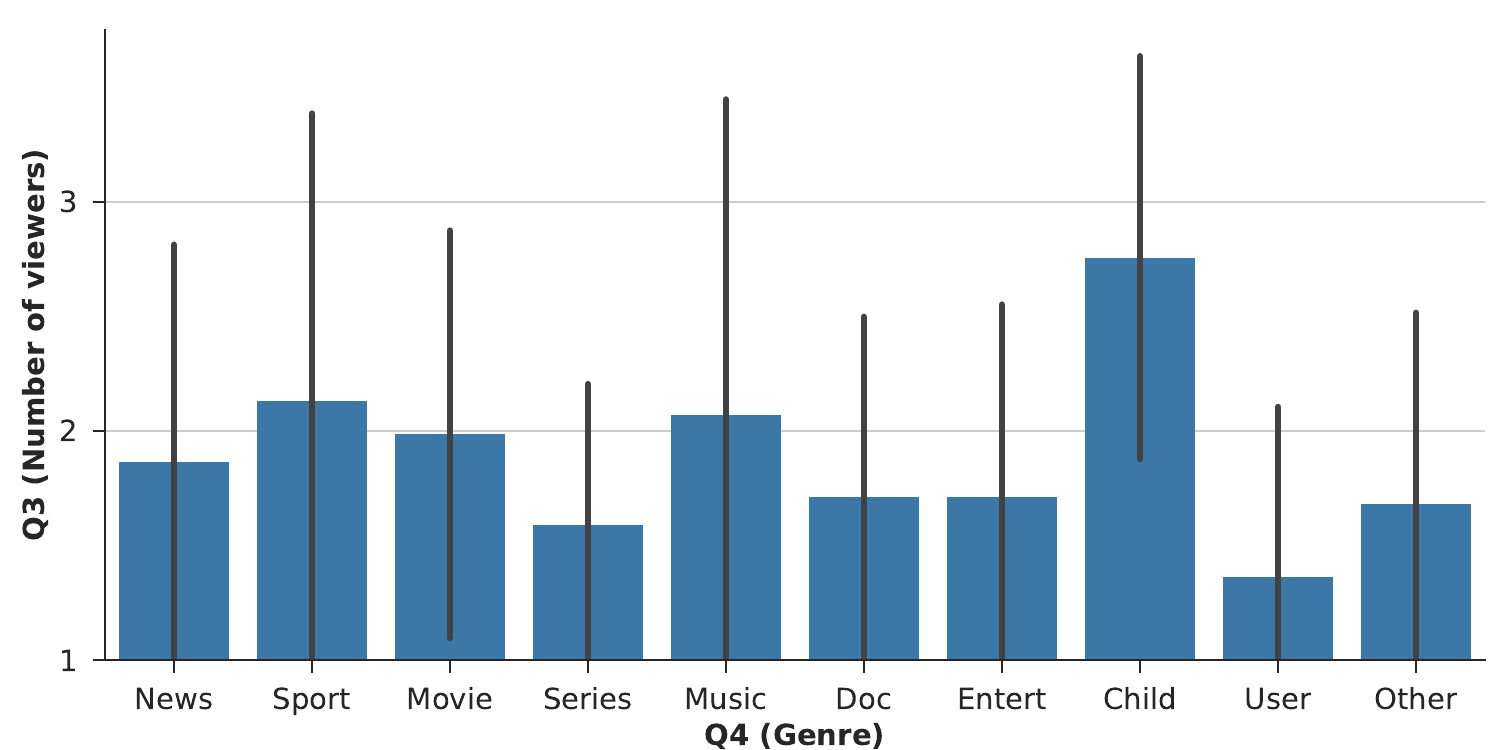}%
    \caption{Top: Average attention level (Q6) reported for each genre (Q4). Bottom: Average number of viewers (Q3) reported for each genre (Q4). The black bars indicate $\pm$one standard deviation.}%
\label{fig:fig3}%
\end{figure}

The dataset consists of 6443 answers. 
Each answer including more than one selected option for Q4 is split (with the same values for the other entries), which brings the total number of answers to 7201.
From these, 3090 are answers with \textit{yes} for Q1.
Fig.~\ref{fig:fig2} shows the distribution of answers for Q1 and Q4.
It is worth noticing that \textit{series} as the most frequent selection for Q4 accounts for approximately 25\% of the answers.
This is in accordance with the reported TV favorites in the background information of the participants, where 75\% of the participants reported \textit{series} as a favorite.
When comparing the reported TV content favorites with Fig.~\ref{fig:fig2}, two genres stand out in particular, namely \textit{movie} and \textit{documentary}.
These are close competitors for second place among reported TV favorites, but in answers for Q4 they have relatively low counts.
This may not come as a surprise, since they typically have a longer duration and might require more attention than the other genres, and thus may not be consumed as frequently.
In addition to being the most watched genre, on average participants report to be slightly more attentive when watching \textit{series} compared to both \textit{movie} and \textit{documentary}, as shown in the top of Fig.~\ref{fig:fig3}.
The reason for the low counts of \textit{movie} and \textit{documentary} needs more analysis, and it would be interesting to study whether it is because users struggle to find and select content within these two genres in particular.

Another point to highlight in Fig.~\ref{fig:fig3} is that \textit{music} and \textit{children's} have low average attention levels compared to the other genres.
In the case of \textit{children's} it is most likely because the TV is used primarily by children of the respondents.
Also, as seen from the bottom of Fig.~\ref{fig:fig3}, \textit{children's} shows mainly to be a social genre. 
Notice the relatively large standard deviation within the genres \textit{sport} and \textit{music} indicating that these are consumed in different social settings, sometimes by one user and at other times by groups of users.

\begin{table}[b]
\centering%
\caption{Pearson's chi-square tests of association between choice of genre and contextual features}%
\label{tab:chi2}%
\begin{tabular}{lcccc}%
\toprule%
    \textbf{Feature}          & $\pmb \chi^{\pmb 2}$ & \textbf{dof} & \textbf{p} & $\pmb V$\\
\midrule%
    Time of day               & 326.39   &  32  &  $<0.001$  &  0.16\\
    Weekday/[weekend/holiday] & 124.52   &  8   &  $<0.001$  &  0.20\\
    Additional viewers        & 1192.53  &  56  &  $<0.001$  &  0.21\\
    Number of viewers         & 540.47   &  32  &  $<0.001$  &  0.21\\
    Attention level           & 593.36   &  32  &  $<0.001$  &  0.22\\
    Service                   & 2169.12  &  56  &  $<0.001$  &  0.29\\
\bottomrule%
\end{tabular}%
\end{table}%

Pearson's chi-square test is used to measure the level of association between the choice of genre (removing answers with the selection \textit{other}) and the contextual features.
Also, Cram\'er's $V$ is reported.
To this end, a contingency table is formed for each contextual dimension, and results are presented in Table~\ref{tab:chi2} for all cases where at least 80\% of the expected frequencies are above five and none are zero.
Statistical significant interactions are found for all features.

\subsubsection{Temporal and Social Aspects of TV Consumption}
We have shown that the choice of genre is associated with several contextual settings.
However, questions, such as what changes during the day, in the weekend, or in social situations, remain unanswered, and are hence investigated below.

\begin{figure}[tb]%
    \centering%
    \includegraphics[width=1.00\columnwidth]{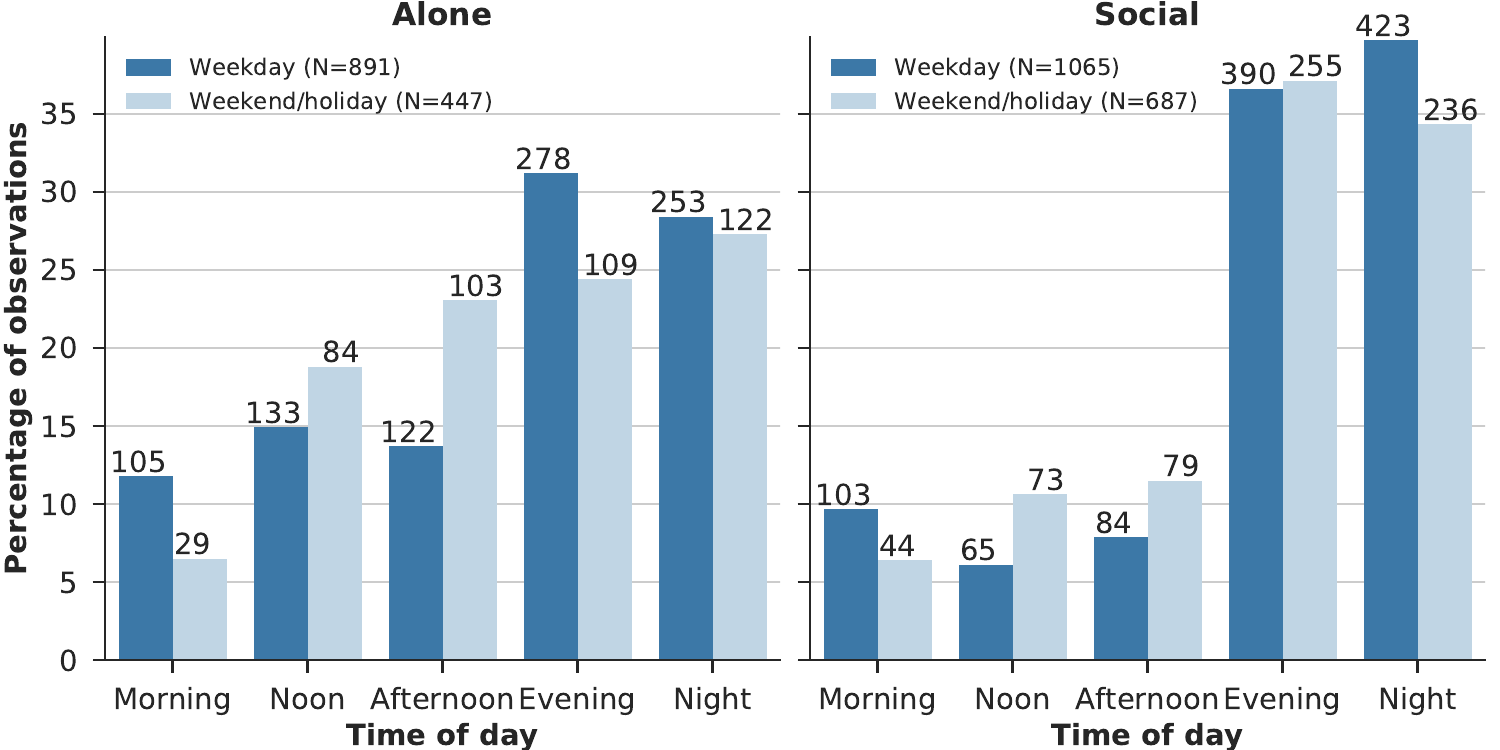}%
    \caption{The distribution of TV consumption according to time of day in four contextual settings, alone/social and weekday/weekend, such that e.g. alone+weekday sums to 100\%. The numbers above the bars indicate the actual number of observations.}%
\label{fig:fig4}%
\end{figure}

A fundamental difference is the consumption pattern.
As shown in Fig.~\ref{fig:fig4}, TV watching in social settings (with at least one co-viewer) happens most frequently during the evening or night both for workdays (75\%) and weekends (70\%).
Morning, noon, and afternoon account for the remaining 25\% of observations for weekdays and 30\% for weekends in social situations.
TV watching in solitary settings is more spread throughout the day, though weekdays show the same tendency with evening and night being the dominant time slots.
Independent of social context, TV watching in the morning occurs most frequently during weekdays, while the share of noon and afternoon viewing increases in weekends.
Also note from the figure that approximately 57\% of all observations (3090) take place in a social context (1752), while 63\% are during workdays (1956).
\begin{figure}[tb]%
    \centering%
    \includegraphics[width=1.00\columnwidth]{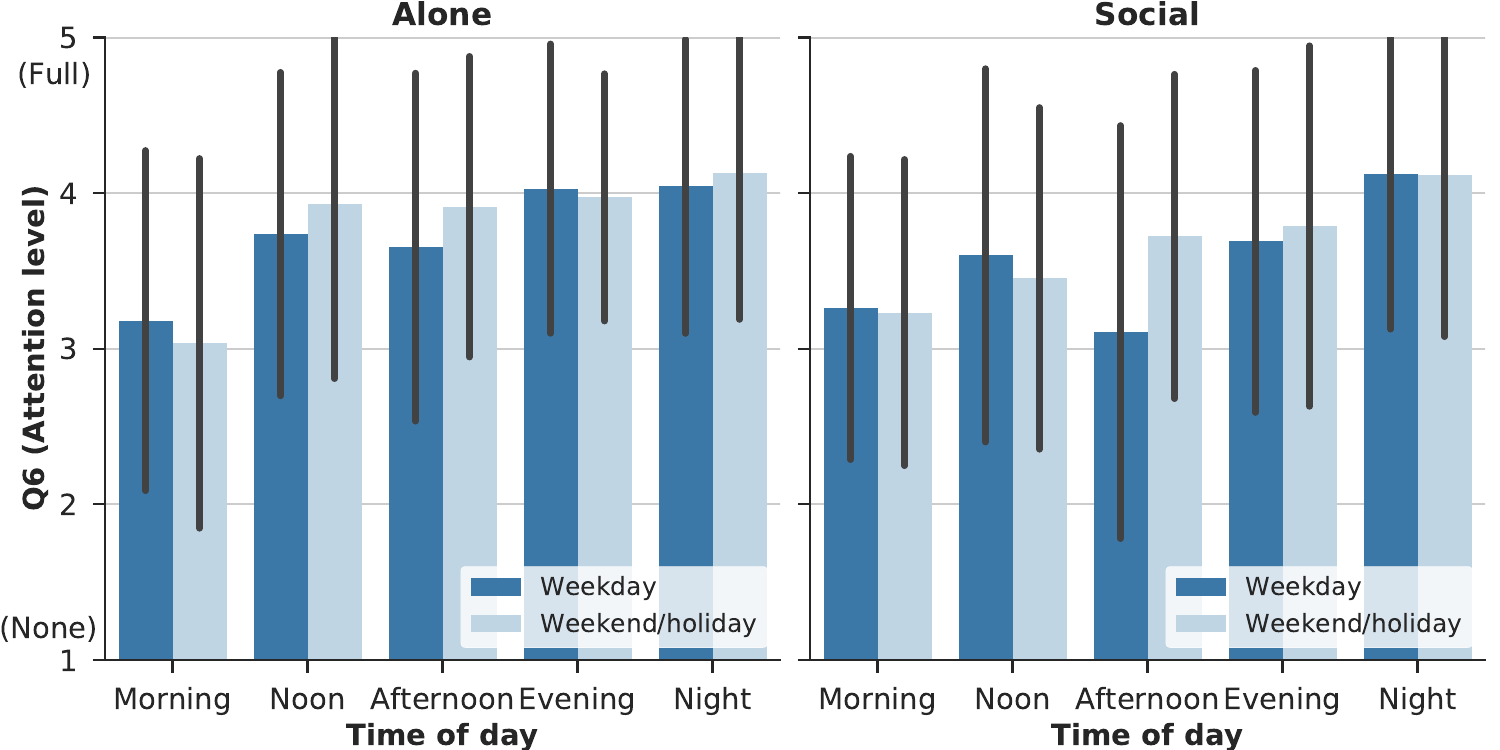}%
    \caption{The average attention level according to time of day in four contextual settings,  alone/social and weekday/weekend. The black bars indicate $\pm$one standard deviation.}%
\label{fig:fig5}%
\end{figure}
\begin{figure*}[t]%
    \centering%
    \includegraphics[width=0.90\textwidth]{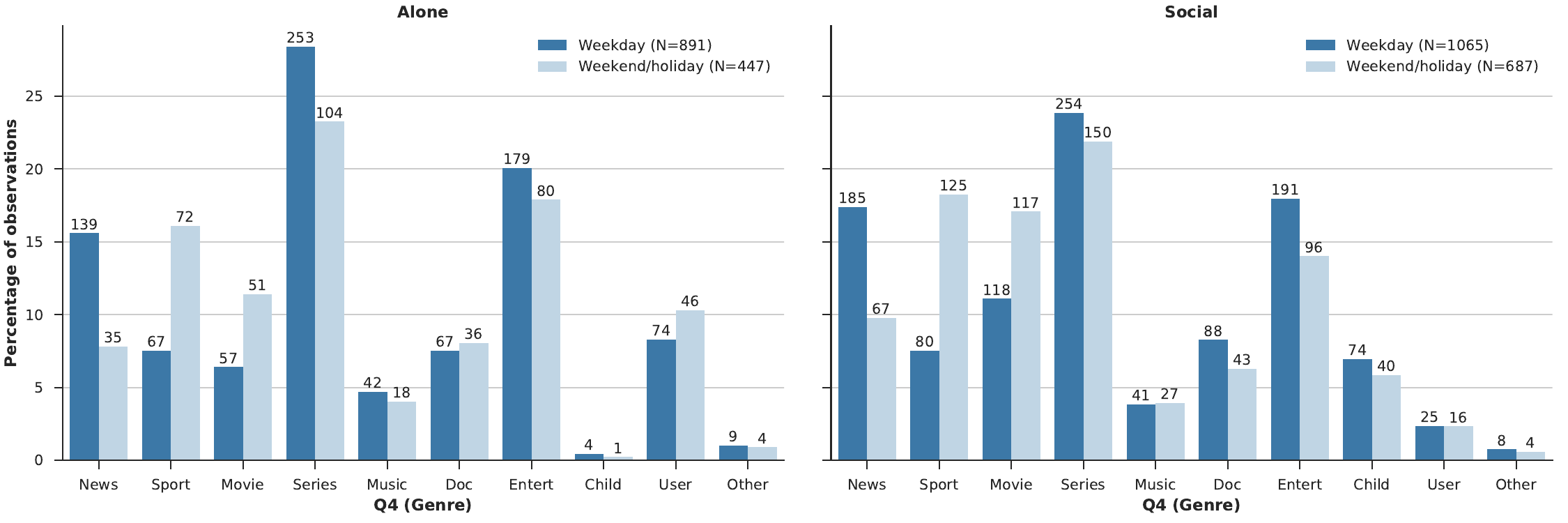}%
    \caption{The distribution of genres consumed in four contextual settings, alone/social and weekday/weekend. The numbers above the bars indicate the actual number of observations.}%
\label{fig:fig6}%
\end{figure*}

Another element to consider is the attention level of the users as presented in Fig.~\ref{fig:fig5}.
Generally, users pay more attention to the content as the day progresses.
A notable exception is social afternoons in weekdays that have the lowest average attention level among all social settings, which could possibly be because users have just returned back home from work and engage in conversations and other activities while watching TV.
There is also a tendency that users pay more attention when they are alone.
Exceptions are mornings and nights, where the levels are approximately similar.
Interestingly, when compared to watching alone, it seems that the way users co-view TV changes between afternoon, evening, and night, such that the social activity tends to be more focused around watching TV intensively as it gets late.

Lastly, Fig.~\ref{fig:fig6} shows how the temporal and social context influence the choice of genre.
Note that the height of the bars indicates the consumption share of a genre within one of the four contextual settings, alone+weekday, alone+weekend, social+weekday, and social+weekend.
Hence, it can for example not be concluded that \textit{movie} is watched more in weekends than weekdays in social settings (actually, the opposite is the case).
It can, however, tell that the proportion of times \textit{movie} is selected over other genres is higher in the weekend.
A number of genres show clear differences between weekdays and weekends.
\textit{News}, \textit{series}, and \textit{entertainment} are preferred during weekdays, while \textit{sport} and \textit{movie} increase their share considerably during weekends.
Though \textit{series} are watched less in the weekends, it is the genre with the largest share for all contexts.
\textit{Entertainment} is second in three out of four contexts, but drops behind \textit{sport} and \textit{movie} to fourth for social weekends.
\textit{News} follows just behind \textit{entertainment} in weekdays, but it is the genre that decreases most in weekends.
\textit{Movie} and \textit{children's} are preferred in social settings, while \textit{user-generated} is mainly consumed when alone, which could possibly be because the genre has its main roots on smaller screens, where users mainly consume it in solitary settings.
The proportion for \textit{music} is similar among the four contexts.

\section{Prediction of Preferences}\label{sec:pred}
In this section we present how the contextual features of the CTV dataset can be used for prediction of consumed content.
The rationale behind is that not only will a user's viewing history influence the choice, but so will the circumstances in which that user is watching TV, which is motivated by the findings of the previous section that shows how contextual settings influence viewing situations contained in the collected dataset.
The procedure and experiments presented in this section provide insights into the mapping between multimedia content genres available on modern TVs and users' personal preferences, and how this mapping can be coupled with context information.

The goal of the methods described in this section is to predict what genre a user is going to watch in the reported context.
Specifically, let $\mathcal{Y}$ be the set of possible genres to choose from in Q4 (see Table~\ref{tab:ques}).
Then $\hat y$ is the predicted genre using a trained classifier, $f$, with input $\pmb x \in \mathbb{R}^D$ consisting of user and context features:
\begin{equation}
    \hat y = f(\pmb x)\text{, where }f: \mathbb{R}^D \rightarrow \mathcal{Y}.
    \label{eq:pred}
\end{equation}
Various methods can be applied to this challenge, such as state-of-the-art factorization machines \cite{Rendle2011} or neural networks \cite{Cheng2016}.
The focus of this investigation is on the contribution of the contextual dimensions and the type of errors, for which reason we do not present the results of a wide palette of algorithms, but rely on a few that are well-known within the machine learning community for establishing baselines of the CTV dataset.\footnote{The dataset includes code for the baselines presented in this section.}
We do, however, show how temporal and social context affect the prediction ability, and compare it to e.g. contextless prediction.

\subsection{Features and Methods}
The task, as shown in \eqref{eq:pred}, is defined as a multiclass classification problem with the users' selections for Q4 as target.
The selections for the remaining questions are used as contextual features (see Table~\ref{tab:feat}).
All features are categorical and represented using one-hot encoding.
Neither user demographics, nor the optional text input for \textit{other} in Q2, Q4, and Q5 are included in this study.

\begin{table}[t]
\centering
\caption{Features and examples of feature configurations}
\label{tab:feat}
\begin{tabular}{p{0.03\columnwidth}lcc}
\toprule
    \multicolumn{2}{l}{\textbf{Feature}}     & \textbf{Dimensions} & \textbf{Origin}\\\midrule
    (\textit{U}) & User ID                   & 118                 & Login\\
    (\textit{T}) & Time of day               & 5                   & Timestamp\\
    (\textit{D}) & Weekday/weekend           & 2                   & Timestamp\\
    (\textit{W}) & Additional viewers        & 8                   & Q2\\
    (\textit{M}) & Number of viewers         & 5                   & Q3\\
    (\textit{S}) & Service                   & 8                   & Q5\\
    (\textit{A}) & Attention level           & 5                   & Q6\\
\midrule
    \multicolumn{4}{l}{\textbf{Feature Configurations}}\\\midrule
    \multicolumn{2}{l}{\textit{all}}         & 143                 & \textit{UTDWMA}\\
    \multicolumn{2}{l}{\textit{all+S}}       & 151                 & \textit{UTDWMSA}\\
    \multicolumn{2}{l}{\textit{all-U}}       & 25                  & \textit{TDWMA}\\
\bottomrule
\end{tabular}
\end{table}

Table~\ref{tab:feat} also lists a number of feature configurations.
In this work, we define the feature configuration \textit{all} as service-independent.
Thus, \textit{all} is a collection of all features except the \textit{service} feature.
The reasoning behind this definition is that a prediction or recommendation of genre will have most impact across providers, since some services have a very targeted range of content genres, e.g. YouTube relies heavily on the user-generated genre.
The \textit{service} feature is included in the \textit{all+S} configuration.
Other configurations can be used as well, e.g. time of day and weekday/weekend (\textit{TD}).
A notable configuration is what we refer to as the contextless, which consists of purely user identity information (\textit{U}).

We include six approaches in the experiment to establish baselines for the CTV dataset.
The methods are selected based on their well-documented achievements in numerous domains.
Four methods are compared using scikit-learn~\cite{Pedregosa2011} implementations: logistic regression (LR), gradient boosting decision trees (GBDT), support vector machines (SVM), multi-layer perceptrons (MLP).
Additionally, we include the most popular (toppop) and random (random) predictors.
For toppop, genres are ranked by their popularity judged by the number of observations in the training set.
The random predictor randomly ranks the genres for each prediction.

\subsection{Evaluation}
The methods are evaluated using nested cross-validation (also referred to as double cross-validation) with five outer folds and three inner folds.
That is, in the outer loop the dataset is split into five folds, using one fold in turn as test set and the remaining four folds as training set.
The training set for each outer iteration is further divided into three inner folds for optimization of hyperparameters.
To this end, on a rotational basis, two (inner) folds are used for training and one is used as validation set.
The best scoring hyperparameter configuration of the inner loop is used to assess the predictive performance of the model on the outer test fold by training on the full training set.
We report the average performance across the outer folds and the standard deviation.
Users that had not answered at least five times were not included in the evaluation.

\subsubsection{Configuration of Hyperparameters}
Due to considerations of computational complexity, some hyperparameters are determined empirically prior to run-time and are static throughout feature configurations.
We fit the LR weights using stochastic average gradient descent with L2 regularization, and set the multi-class parameter to ''multinomial'' for softmax regression.
For GBDT, we use 1000 boosting stages, each fitted on a random subsample consisting of 50\% of the training samples.
The SVM use a one-vs-rest decision scheme, and MLP is implemented with two hidden layers each consisting of 200 neurons with rectified linear (ReLU) activation functions and a softmax output layer, optimized during training using Adam and L2 regularization of weights.
During hyperparameter tuning in the inner loop of the nested cross-validation, the following variables are determined: LR - regularization strength; GBDT - maximum depth of individual trees; SVM - kernel type (linear/RBF), kernel coefficient and regularization strength; MLP - regularization strength.

\subsubsection{Metrics}
The hit ratio at K predictions (HR@K) is used as a metric for evaluation. 
At K larger than one, multiple guesses are allowed for each trial.
It is calculated using:
\begin{equation}
    \text{HR@K} = \frac{1}{N}\sum_{n=1}^{N}\sum_{k=1}^{K}\pmb 1 \left( \hat y_{n,k} = y_n \right)
\end{equation}
where $N$ is the number of trials. $\pmb 1$ is the hit/indicator function, which is one if the prediction, $\hat y_{n,k}$, is equal to the actual target, $y_n$, and zero otherwise.
The $\hat y_{n}$ is sorted with predictions in ascending order according to confidence score, such that $\hat y_{n,1}$ is the most probable prediction.
The mean reciprocal rank (MRR) is used to assess the average ranking ($k$) of the true targets:
\begin{equation}
    \text{MRR} = \frac{1}{N}\sum_{n=1}^{N}\sum_{k=1}^{M}\frac{1}{k}\pmb 1 \left( \hat y_{n,k} = y_n \right)
\end{equation}
where $M$ is the total number of target classes.
We also report F1 scores with macro averaging.
F1 (macro) is an average of each individual class performance.
It gives equal weight to classes, which means it can be used to assess the performance on small classes.
This makes F1 (macro) an indicator of a method's ability to predict diverse target classes.
On the other hand, F1 (micro) pools all trials with equal weight. 
Thus, classes with many samples will dominate classes with few samples. 
F1 (micro) is with this setup (multi-class single-label) equal to HR@1 and therefore not presented explicitly.

\subsection{Results}\label{sec:res}
\subsubsection{Feature configuration \textit{all}}
The performance using feature configuration \textit{all} is shown in Table~\ref{tab:res}.
It is not a surprise that random scores approximately 0.1 in HR@1 (10 target classes), and that toppop scores close to 0.25 since it was shown that \textit{series} as the most watched genre accounts for approximately one fourth of the data points.
Also, toppop outperforms random in terms of HR@3 and MRR.
Note, however, that random performs better than toppop for F1 (macro), due to the diversity in predicted genres. 

The remaining methods achieve considerably higher scores than random and toppop.
LR, as the best scoring, almost doubles the HR@1 of toppop, and successfully predicts the genre in approximately 44\% of the cases and 82\% when allowing three guesses.

\begin{table}[b]
\centering%
\caption{Results for the feature configuration \textit{all} \protect\\ (standard deviation in parentheses)}%
\label{tab:res}%
\resizebox{\columnwidth}{!}{%
\begin{tabular}{ccccc}%
\toprule%
\textbf{Method} & \textbf{HR@1}  & \textbf{HR@3}  & \textbf{F1 (macro)} & \textbf{MRR}\\
\midrule
random          & 0.101 (0.005) & 0.295 (0.004) & 0.087 (0.008) & 0.290 (0.005)\\
toppop          & 0.245 (0.009) & 0.560 (0.014) & 0.039 (0.001) & 0.460 (0.008)\\
MLP             & 0.413 (0.018) & 0.787 (0.013) & 0.335 (0.022) & 0.621 (0.012)\\
GBDT            & 0.417 (0.014) & 0.786 (0.021) & 0.354 (0.027) & 0.623 (0.009)\\
SVM             & 0.425 (0.009) & 0.809 (0.021) & 0.358 (0.015) & 0.632 (0.012)\\
LR              & 0.437 (0.026) & 0.815 (0.010) & 0.373 (0.031) & 0.641 (0.013)\\
\bottomrule%
\end{tabular}%
}%
\end{table}%

\subsubsection{Genre confusions}
\begin{figure}[t]%
    \centering%
    \includegraphics[width=1.0\columnwidth]{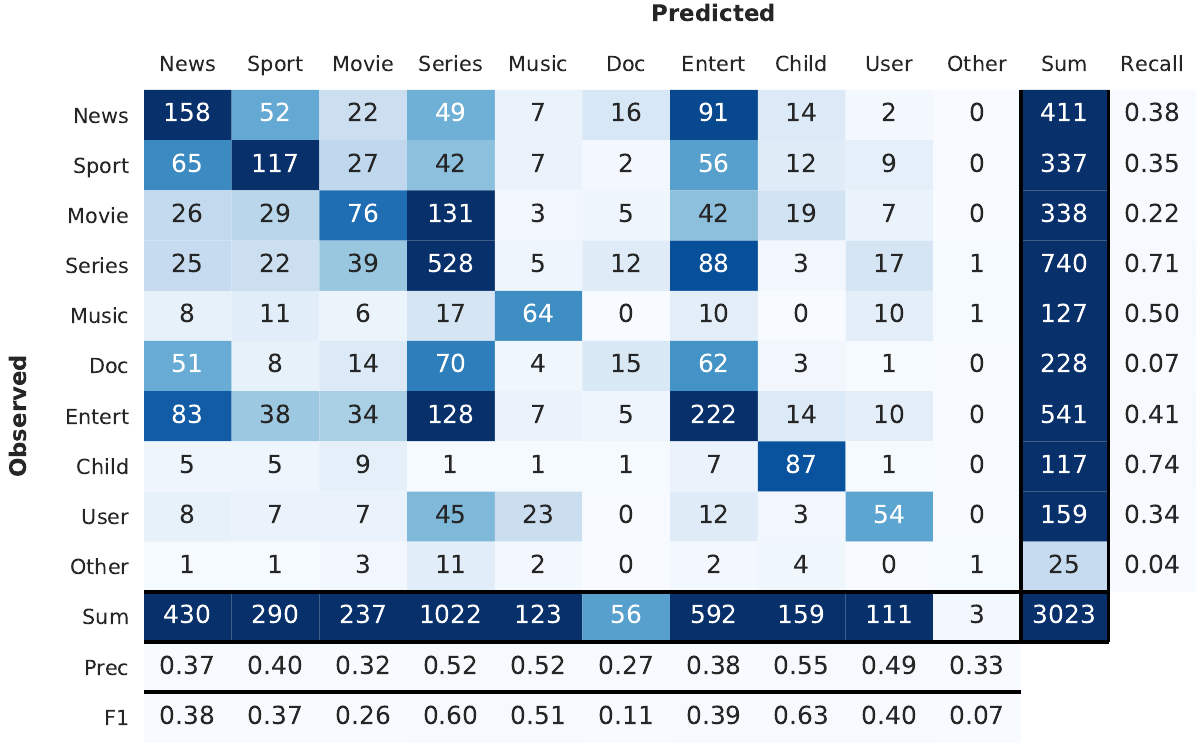}%
    \caption{Confusion matrix using LR for feature configuration \textit{all} and the aggregated (out-of-fold) cross-validation predictions. The bottom rows show the precision and F1 for each genre, and the right-most column shows the recall.}%
\label{fig:fig7}%
\end{figure}

A deeper look into the prediction errors of LR for feature configuration \textit{all}  is shown in a confusion matrix in Fig.~\ref{fig:fig7}.
\textit{Series} receives many predictions (1022) compared to the actual number of occurrences (740) resulting in the best recall score of all the genres (true positives over number of observations, $528/740=0.71$).
Despite the many predictions, it also manages to achieve one of the highest precision scores (true positives over number of predictions, $528/1022=0.52$).
The opposite is the case with \textit{documentary}, which receives few predictions (56) compared to observations (228), leading to a low recall score (0.07).
Together with the precision score of 0.27, the resulting F1 score is 0.11, which emphasizes that \textit{documentary} is difficult to predict in this specific setup.
It can be seen from the off-diagonal entries that there are three main genres that are confused with \textit{documentary}, being \textit{news}, \textit{series}, and \textit{entertainment}.

Two genres stand out in terms of F1 score, namely \textit{series} and \textit{children's} that both achieve a score around 0.6.
Next is \textit{music} with a score of 0.5, which gets a majority of its false positives when the observed genre is \textit{user-generated}.
The \textit{user-generated} genre has a high (relative) precision, but a low recall causes the F1 score to fall to an average level among the genres, meaning that even though it is hard to retrieve \textit{user-generated} content for recommendation, when it is finally predicted, the predictions are fairly reliable.
\textit{News}, \textit{sport}, and \textit{entertainment} also have F1 scores that are close to the F1 (macro).
In addition to \textit{documentary}, the \textit{movie} genre underperforms.
The F1 score of \textit{movie} is low mainly due to a low recall, caused to a high degree by \textit{series}, but it is also frequently confused with \textit{news}, \textit{sport}, and \textit{entertainment}.

\subsubsection{Contextual dimensions}
\begin{figure*}[tb]%
    \centering%
    \includegraphics[width=0.49\textwidth]{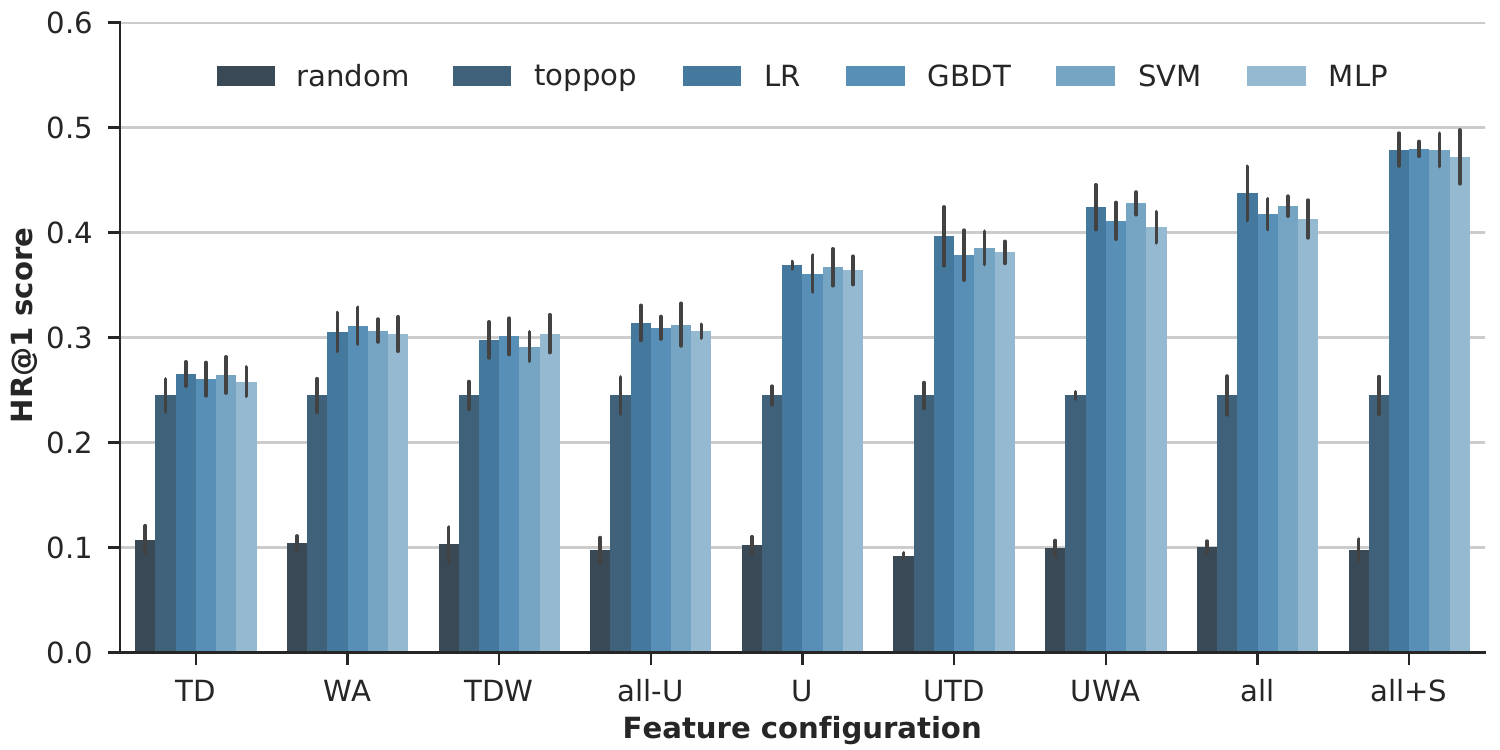}%
    \hfill
    \includegraphics[width=0.49\textwidth]{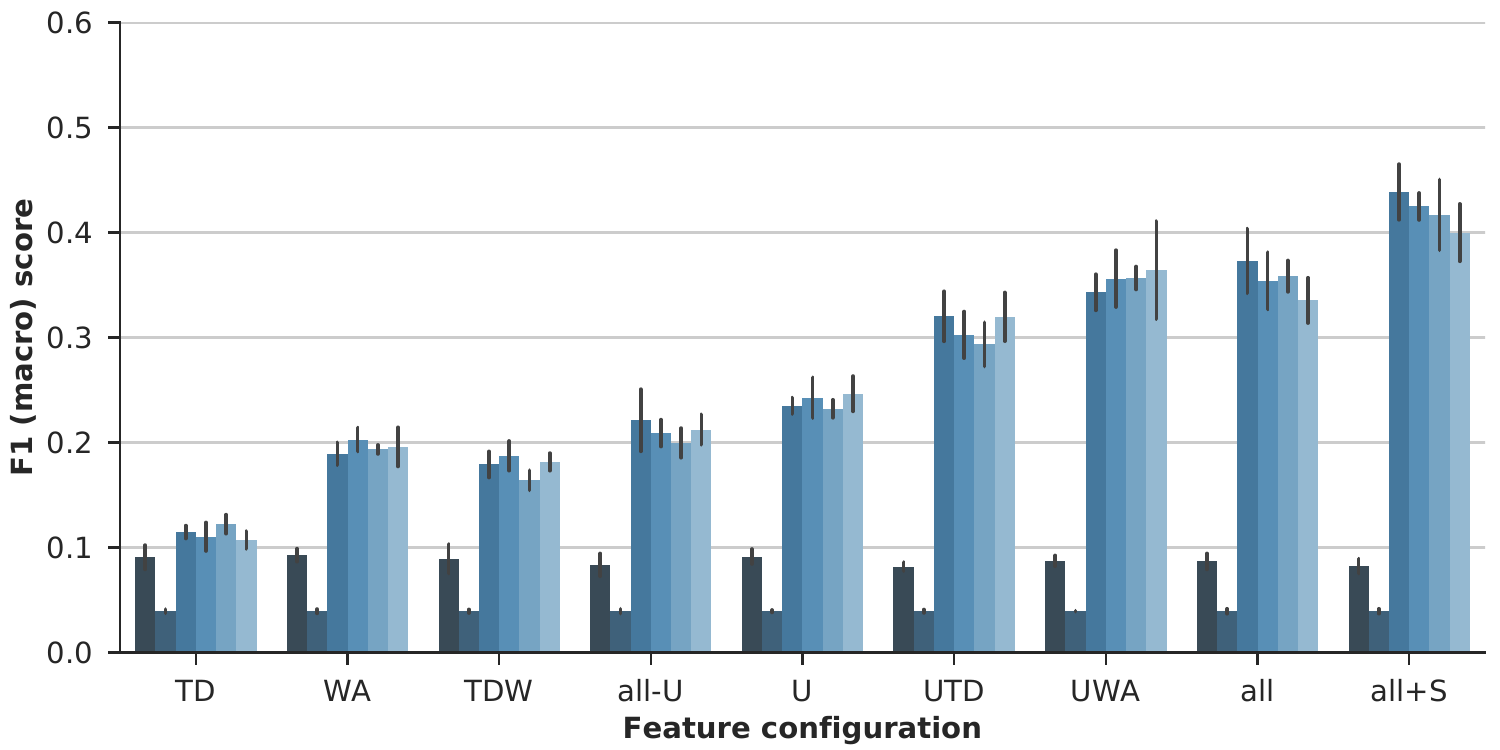}\\%
    \vspace{1em}%
    \includegraphics[width=0.49\textwidth]{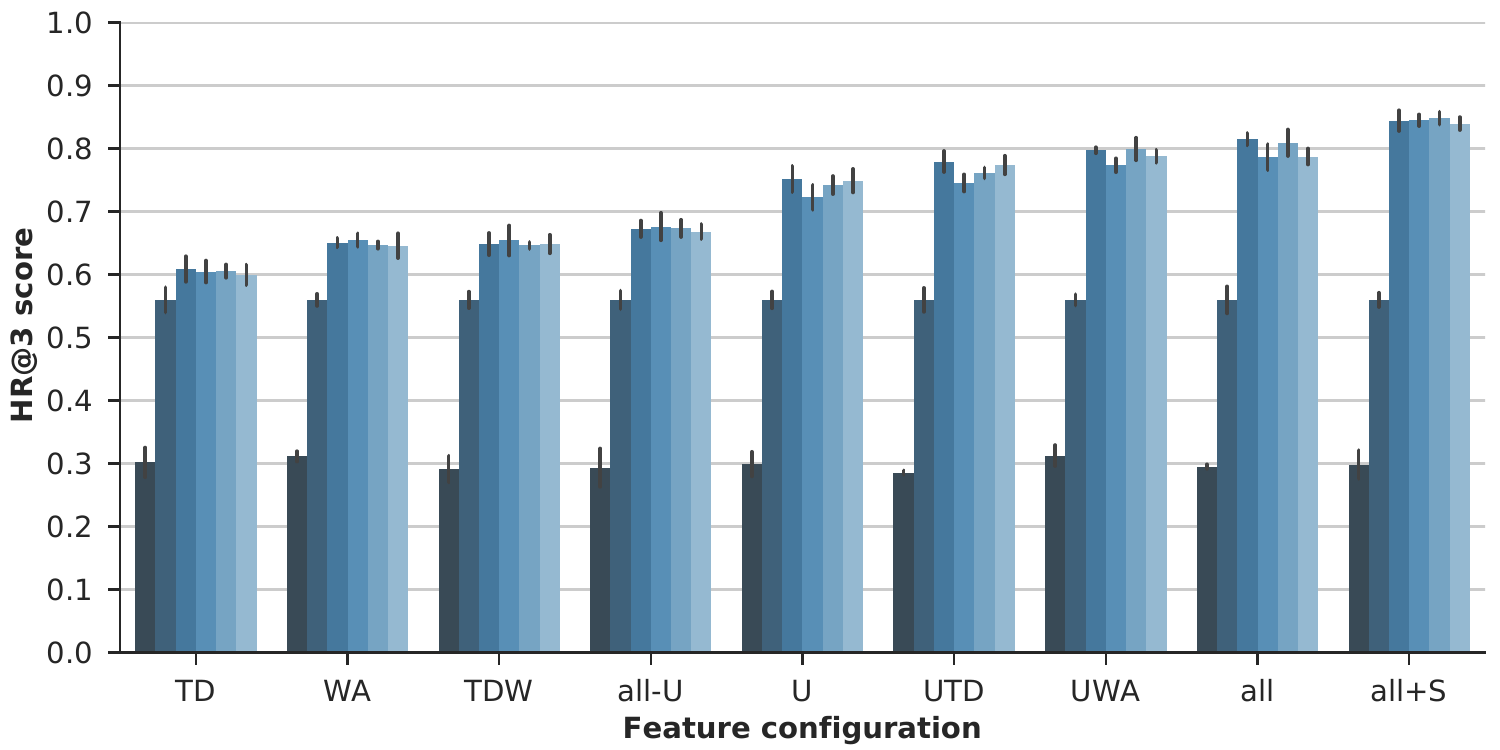}%
    \hfill
    \includegraphics[width=0.49\textwidth]{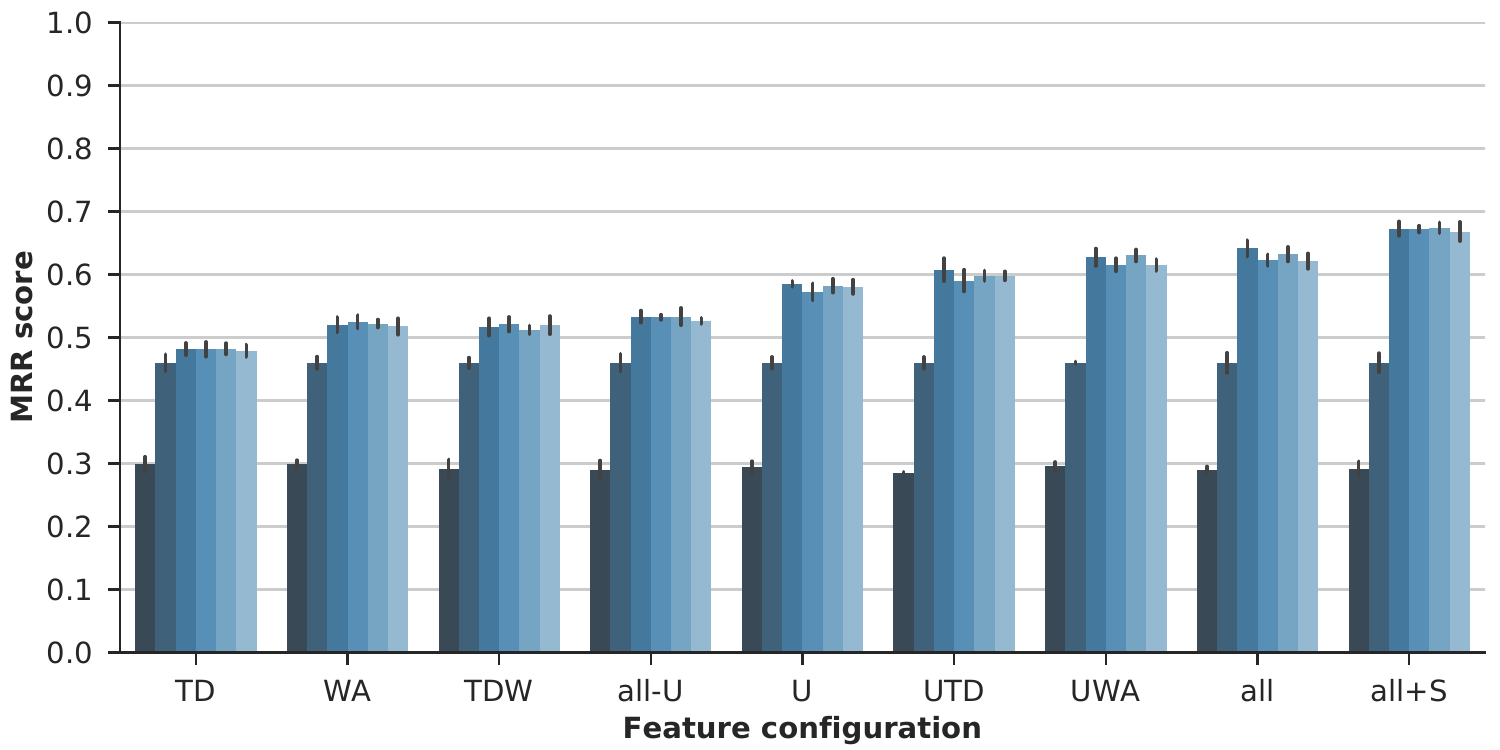}%
    \caption{HR@1, HR@3, F1 (macro), and MRR scores for the six methods using different feature configurations (see Table~\ref{tab:feat}). The height of the bars indicate the average of the cross-validation results and the black bars show $\pm$one standard deviation.}%
\label{fig:fig8}%
\end{figure*}

A comparison of the methods for multiple feature configurations is shown in Fig.~\ref{fig:fig8}.
Note that the random and toppop scores are independent of feature selection.
Also, as can be seen from the figure, in general there is not a large deviation between performance of the four top performing methods within each feature configuration.
Therefore, we mainly highlight results between feature configurations in the following.

The worst performing feature configuration in Fig.~\ref{fig:fig8} is the temporal configuration, \textit{TD}, for which the performance of LR almost decreases to comparable results of random and toppop, though the HR@1, HR@3, and MRR are significant better than those of the random predictor, and likewise the F1 (macro) score is higher than what toppop achieves.
Adding social aspects (\textit{TD}$\rightarrow$\textit{TDW}) significantly improves performance according to McNemar's test\footnote{A matrix, $\pmb A_{2 \times 2}$, is formed with $a_{1,1}$ being the number of trials where both methods are correct, $a_{1,2}$ and $a_{2,1}$ are the trials where one of the methods fail, and $a_{2,2}$ is when both are incorrect. McNemar's $\chi^2$ test statistic and Cram\'er's $V$ are then computed as:\\$\chi^2 = \left( a_{1,2} - a_{2,1} \right) ^2 / \left(  a_{1,2} + a_{2,1} \right)$, $V = \sqrt{ \chi^2 / \sum_{i,j}a_{i,j}}$.} (for LR: $\chi^2$(1)=29.10, p$<$0.001, $V$=0.09).
Furthermore, knowledge of additional viewers and attention level, \textit{WA}, scores on par with \textit{TDW}, and even slightly improves in terms of F1 (macro).

Notably, contextless prediction, \textit{U}, outperforms \textit{all-U}, which indicates the importance of user specific behavior and habits (for LR: $\chi^2$(1)=26.33, p$<$0.001, $V$=0.09).
The F1 (macro), however, does not see as large an improvement between the two configurations as the three other metrics.
Adding temporal context to the user ID (\textit{U}$\rightarrow$\textit{UTD}) increases performance slightly, while F1 (macro) improves considerably suggesting that the temporal information enhances the ability to predict diverse genres.
As is the case with \textit{TD}$\rightarrow$\textit{WA}, \textit{UWA} performs significantly better than \textit{UTD} (for LR: $\chi^2$(1)=9.83, p$\approx$0.002, $V$=0.06).
In fact, \textit{UWA} provides results on the same level as \textit{all} with no significant difference according to McNemar's test (for LR: $\chi^2$(1)=3.52, p$>$0.05, $V$=0.03). 
Lastly, knowing which service a user is using, \textit{all+S}, achieves the highest scores.

The contribution of contextual dimensions to the overall result is investigated in Table~\ref{tab:relperf}, which shows the average performance when each variable is removed from the LR model in turn.
The results are relative to LR for the feature configuration \textit{all}.
Again, user identity (\textit{U}) shows to constitute a vital component of the model with substantial performance reductions compared to the other features.
On the contrary, the weekday/weekend feature (D) contributes the least among the inputs.
Interestingly, the metrics are affected differently for the remaining features. 
Removing information of additional viewers (\textit{W}) reduces HR@1 the most, while the time of day (\textit{T}) has the largest impact on F1 (macro).
This indicates distinct strengths among the features.
For example, the time of day helps predict a diverse range of genres, which is also suggested by the previously discussed \textit{U}$\rightarrow$\textit{UTD} results of Fig.~\ref{fig:fig8}.
\begin{table}[tb]
\centering%
\caption{Relative performance when each contextual variable is removed compared to feature configuration \textit{all} using LR}%
\label{tab:relperf}%
\resizebox{\columnwidth}{!}{%
\begin{tabular}{lccccccc}%
\toprule%
    \textbf{Feat. Conf.}    & all-D & all-M & all-T & all-A & all-W & all-U\\
    \textbf{Rel. HR@1}       & -0.9\%& -1.4\%& -2.1\%& -2.3\%& -2.5\%& -29.5\%\\
    \textbf{Rel. F1 (macro)}& -2.4\%& -3.8\%& -5.9\%& -5.1\%& -4.0\%& -44.2\%\\
\bottomrule%
\end{tabular}}%
\end{table}%

\section{Discussion}\label{sec:disc}
The predictive performance using LR for feature configuration \textit{all} (see Fig.~\ref{fig:fig7}) is best for the two genres \textit{series} and \textit{children's} in terms of F1 score.
According to the data analysis (presented in Section~\ref{sec:data_analysis}), \textit{children's} exhibit distinctive contextual preferences, such as low attention level and multiple viewers, resulting in fewer prediction confusions.
The opposite is the case with \textit{documentary} and \textit{movie} that are not easily distinguishable from the features collected in this study, which also makes them the genres with the lowest F1 scores.
A reason for the worse performance when compared to \textit{series}, could be the imbalance of the dataset, and the overall score of the less viewed genres could possibly improve by handling this.
Also, one should keep genre ambiguities in mind when evaluating the system. 
For example, a movie could be a children's movie, thereby potentially covering two genres in this study.
Assuming that a participant provides consistent labels throughout the study, these ambiguities would primarily result in reduced performance when making use of information across users.

The analysis of the collected data suggests that contextual aspects are an integral part of users' decision process when selecting what content to watch on TV.
This hypothesis is supported by the findings of Section~\ref{sec:pred}, in which it is shown that the inclusion of contextual information is positively associated with improved predictive performance, both in terms of accuracy and diversity.
Note that this is without inducing prior expert knowledge of certain situations, such as the often easily distinguishable \textit{children's} setting.
As previously mentioned, user demographics are not included in the experiments, but they can potentially enhance the system further.
However, brief experiments showed that information of age group and gender did not introduce measurable performance improvements of LR when user identity is also included.
That being noted, the results of Section~\ref{sec:pred} also highlight the importance of knowing who is watching, since \textit{all-U}$\rightarrow$\textit{U} shows significant improvement.
Interestingly, though adding temporal information to the user ID (\textit{U}$\rightarrow$\textit{UTD}) only shows slight improvement in accuracy, it clearly indicates more diverse predictions.
Hence, contextless prediction (\textit{U}) achieves high accuracy scores, but it seems to be at the expense of the ability to predict diverse genres.
However, from the results obtained in this work, contextual prediction benefits from knowing user habits, and drawing on contextual patterns of viewing situations should thus not exclude personalization based on past behavior of each individual.

As opposed to the findings of \cite{Cremonesi2015} that the effects of social context are canceled when also considering temporal context, our results suggest a significant improvement when adding social context to temporal, e.g. \textit{TD}$\rightarrow$\textit{TDW}.
This is also evident in terms of diversity in the predicted genres.
What users watch, when, and with whom may be correlated, shown by the habitual behavior of users when watching TV, but our results indicate that knowing the social context of a viewing situation will enable the system to adapt to some scenarios deviating from temporal habits.
As an example, content chosen for Friday nights could be very different depending on the social company of the user, while Monday mornings frequently consist of the same users and content.

\section{Concluding Remarks}\label{sec:conc}
In this paper, we introduced the novel and publicly available CTV dataset of TV consumption enriched with contextual information that contributes to the evaluation of TV viewing in the home.
To this end, we conducted an extensive field study over a period of five weeks with a group of more than 100 participants. 
Using the dataset, we showed associations between different aspects of TV watching, e.g. users' average attention level of genres, and how these change over time and in different social situations.
Furthermore, we evaluated to which degree contextual knowledge influences the performance of predicting what content will be consumed.
The experimental results showed that inclusion of contextual information significantly improves accuracy and diversity compared to contextless predictions, but also that knowledge of past behavior is essential to achieve high accuracies.

In future work we plan to apply methods that have proven successful within the context-aware recommender systems community, and evaluate how they adapt to a dataset with a limited number of target classes and multiple interactions between the same user and target.
This will among others include deep models suitable for inferring latent contextual features.
The presented work is directly applicable to recommending genres.
Future work can potentially investigate using genres for reducing the search space in item recommendations.


\enlargethispage{-9cm}

\begin{IEEEbiography}[{\includegraphics[width=1in,height=1.25in,clip,keepaspectratio]{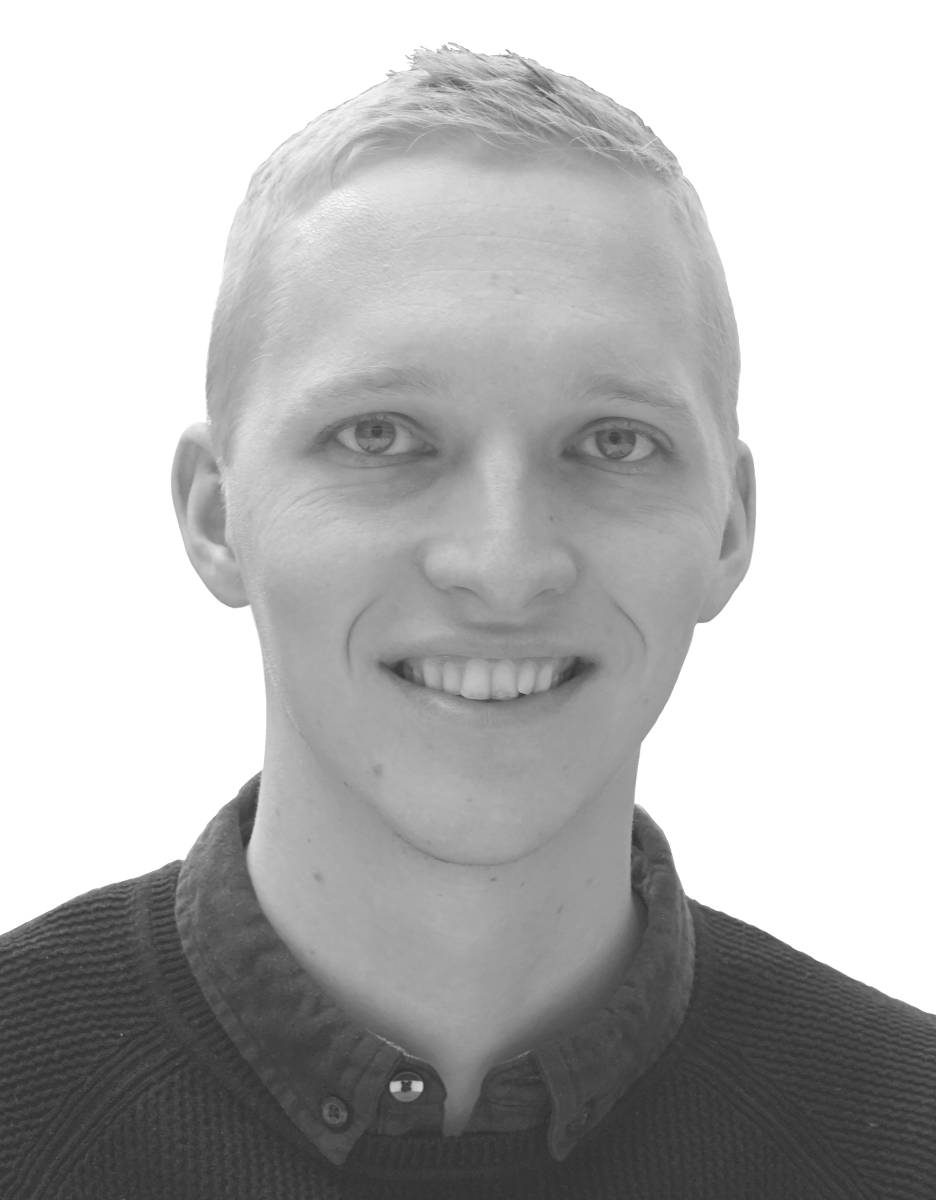}}]{Miklas Str{\o}m Kristoffersen}
(StM'17)
    received the B.Sc. and M.Sc. degrees in electronic engineering with specialization in vision, graphics, and interactive systems from Aalborg University, Aalborg, Denmark, in 2014 and 2016, respectively, where he now pursues the Ph.D. degree. He is currently a research fellow with Bang \& Olufsen A/S, Struer, Denmark. His research interests include computer vision, machine learning, and context-aware recommender systems.
\end{IEEEbiography}

\begin{IEEEbiography}[{\includegraphics[width=1in,height=1.25in,clip,keepaspectratio]{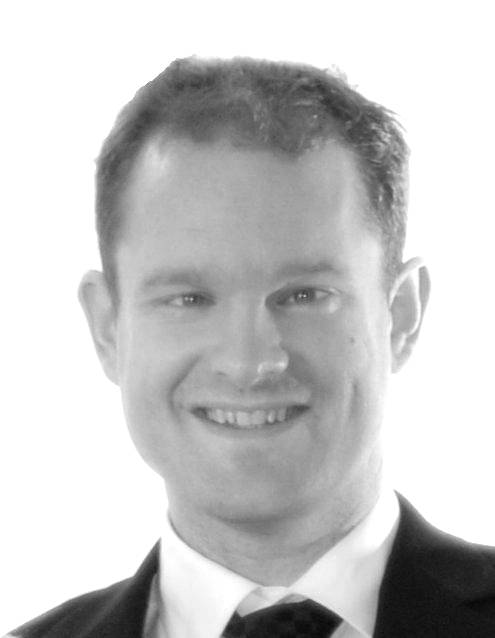}}]{Sven Ewan Shepstone}
(M'11)
received the B.S. and M.S. degrees in electrical engineering from the University of Cape Town, Cape Town, South Africa, in 1999 and 2002, respectively, and the Ph.D. degree from Aalborg University, Aalborg, Denmark, in 2015.

From 2005 to 2010, he researched in the field of broadband communications with Ericsson a/s, Copenhagen, Denmark, and has been with Bang \& Olufsen A/S, Struer, Denmark, since 2010, where he is currently a Research Specialist. His current research interest includes AI applied to consumer electronics. Dr. Shepstone was a recipient of the IEEE Ganesh N. Ramaswamy
Memorial Student Grant at ICASSP 2015.
\end{IEEEbiography}

\begin{IEEEbiography}[{\includegraphics[width=1in,height=1.25in,clip,keepaspectratio]{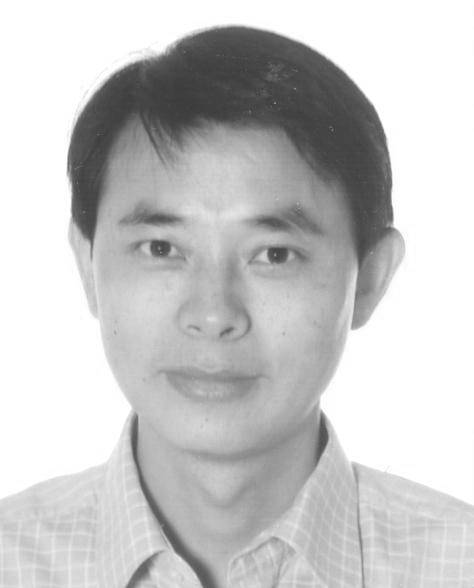}}]{Zheng-Hua Tan}

(M’00–SM’06)
received the B.Sc. and M.Sc. degrees in electrical engineering from Hunan University, Changsha, China, in 1990 and 1996, respectively, and the Ph.D. degree in electronic engineering from Shanghai Jiao Tong University, Shanghai (SJTU), China, in 1999.

He is a Professor in the Department of Electronic Systems and a Co-Head of the Centre for Acoustic Signal Processing Research (CASPR) at Aalborg University, Aalborg, Denmark. He was a Visiting Scientist at the Computer Science and Artificial Intelligence Laboratory (CSAIL), Massachusetts Institute of Technology (MIT), Cambridge, USA, an Associate Professor in the Department of Electronic Engineering at SJTU, Shanghai, China, and a postdoctoral fellow in the Department of Computer Science at KAIST, Daejeon, Korea. His research interests include machine learning, deep learning, pattern recognition, speech and speaker recognition, noise-robust speech processing, multimodal signal processing, and social robotics. He has authored/coauthored about 200 publications in refereed journals and conference proceedings. He is a member of the IEEE Signal Processing Society Machine Learning for Signal Processing Technical Committee (MLSP TC). He has served as an Editorial Board Member/Associate Editor for Computer Speech and Language, Digital Signal Processing, and Computers and Electrical Engineering. He was a Lead Guest Editor of the IEEE Journal of Selected Topics in Signal Processing and a Guest Editor of several journals including Neurocomputing. He is the General Chair for IEEE MLSP 2018 and was a Technical Program Co-Chair for IEEE Workshop on Spoken Language Technology (SLT 2016).
\end{IEEEbiography}

\enlargethispage{-8.5cm}

\end{document}